\documentclass[conference]{IEEEtran}
\IEEEoverridecommandlockouts

\usepackage[nocompress]{cite}

\AtBeginDocument{%
  \providecommand\BibTeX{{%
    \normalfont B\kern-0.5em{\scshape i\kern-0.25em b}\kern-0.8em\TeX}}}

\usepackage{subfig}
\usepackage{subcaption}
\usepackage{booktabs}
\usepackage{tabularx}
\usepackage{tikz}
\usepackage{array}
\usepackage{enumitem}
\usepackage{amsmath}
\usepackage[colorlinks,bookmarksopen,bookmarksnumbered,citecolor=red,urlcolor=blue]{hyperref}
\usepackage{comment}
\usepackage[normalem]{ulem}
\usepackage{balance}

\usepackage[dvipsnames,svgnames,x11names]{xcolor}

\begin{document}

\title{Avatar: Toward Autonomous End-to-End Orchestration of Scientific Workflows using LLMs}

\author{
\IEEEauthorblockN{Suman Raj$^{1,2}$, Hai Duc Nguyen$^{2}$, Haochen Pan$^{1}$, Ryan Chard$^{2}$, Kyle Chard$^{1,2}$, Ian Foster$^{2,1}$}
\IEEEauthorblockA{$^{1}$Department of Computer Science, University of Chicago, IL, USA\\
$^{2}$Data Science and Learning Division, Argonne National Laboratory, Lemont, IL, USA\\
Email: \{sumanraj, haochenpan, chard\}@uchicago.edu, \{hai.nguyen, rchard, foster\}@anl.gov}
}

\maketitle
\thispagestyle{plain}
\pagestyle{plain}

\begin{abstract}
Scientific workflow management (WMSs) systems automate execution, yet orchestrate using fixed, hand-tuned rules. LLM agents promise more autonomous orchestration, but it remains unclear where to introduce agentic reasoning, how to bound its risk, and when it actually helps.
We present Avatar, an actor-based architecture comprising an orchestrator, an executor, and a provenance monitor. Each actor's decision policy is pluggable (rule-based or LLM-backed) via a single adapter-validated action catalog, so conventional and agentic control run on the same core across different WMSs. We present an implementation using the Academy framework and evaluate Avatar across three workloads. We observe that Avatar's rule mode reproduces native execution, with a single unchanged core running all three. Moreover, LLM-backed Avatar reports a reduction of compute wastage by $55\%$ and cuts GPU-busy time by $40\%$. Overall, we envision Avatar as a step toward workflow systems that reason about their own orchestration rather than follow pre-fixed rules.

\end{abstract}

\section{Introduction}

Scientific discovery increasingly relies on complex computational campaigns that combine simulation, data analysis, machine learning, distributed instruments, and heterogeneous computing resources~\cite{ferreiradasilva2024frontiers}. At the same time, recent advances in large language models (LLMs) and agentic AI are changing how users interact with computational systems.
This capability is particularly promising for scientific workflows, where researchers often spend substantial effort translating scientific intent into executable workflows, configuring runtime systems, monitoring execution, diagnosing failures, and adapting computation as conditions change\cite{thareja2026specification,balis2026researchquestion}. 
Agentic reasoning offers the potential to automate or assist many of these decisions, moving workflow management toward more adaptive and autonomous orchestration.

Scientific workflow management systems (WMS) such as Pegasus \cite{deelman2015pegasus} and Parsl \cite{babuji2019parsl} already provide many of the abstractions needed for this transition. 
These abstractions transform workflow management into a set of explicit, inspectable, and actionable interfaces through which agents can reason over workflow graphs, revise task plans, select resources, interpret monitoring feedback, and diagnose failures from logs and provenance. Recent efforts make this direction more visible by extending workflow systems toward decentralized decision making, hybrid AI-HPC execution, federated agents, and AI-assisted workflow lifecycle management \cite{alsaadi2025rhapsody,pauloski2025academy,thareja2026specification}. Together, these developments indicate a broader shift toward workflow systems in which AI agents can participate directly in orchestration rather than remaining external assistants.

This transition, however, exposes a fundamental architectural gap. Most existing WMSs were designed around conventional algorithms, policies, and human-configured control rather than agentic reasoning as a first-class capability. Adding agents therefore commonly requires retrofitting them into systems whose abstractions, control boundaries, and runtime interfaces were designed for different assumptions.
Such adaptation is particularly challenging because mature scientific WMSs are large, well-established, and operationally complex, making substantial architectural changes costly and time-consuming~\cite{shin2025revolution}.
More importantly, without a common understanding of where agents should be introduced, what information they should observe, which decisions they should control, and how their actions should be validated, each system must resolve these questions independently. The resulting integrations can consequently become narrowly scoped, difficult to generalize, or expensive to evaluate across different models, workflows, agent configurations, and infrastructure conditions.

We address this gap with \textbf{Avatar}, a reference architecture for agentic scientific workflow orchestration. Rather than organizing the architecture around the capabilities of a particular LLM, agent framework, or WMS, Avatar is derived from recurring responsibilities across existing scientific workflow systems.
It distills these responsibilities into three interacting components: an orchestrator that manages logical workflow decisions, an executor that realizes those decisions on runtime resources, and provenance that observes and maintains the evidence needed for subsequent decisions. Each component exposes explicit state, decision, and action interfaces through which conventional policies, LLM-backed agents, or hybrid approaches can be introduced without changing the surrounding architecture.
This separation provides a common experimental basis for studying where agentic reasoning should be placed, what responsibilities it should assume, and how it interacts with conventional workflow mechanisms. By holding the architecture fixed while varying the reasoning policy, Avatar enables controlled evaluation of when agentic orchestration helps, when conventional mechanisms remain preferable, and what architectural factors determine the outcome.
Specifically, the contributions of the paper are as follows: 
\begin{itemize}[leftmargin=*]
\item We propose \textit{Avatar}, a reference architecture for agentic scientific workflow orchestration. The architecture captures common components of existing scientific workflow frameworks and exposes flexible integration points for adding LLM-backed reasoning capability (\S\ref{sec:reference-arch}).

\item We implement \textit{Avatar} as an extensible prototype framework using Academy, allowing different agents, tools, monitoring interfaces, and validation strategies to be integrated and compared systematically (\S\ref{sec:implementation}).

\item We evaluate \textit{Avatar} across diverse workloads, orchestration structures, and objectives, revealing when agentic reasoning can reduce wasted computation by up to $55\%$ and GPU-busy time by $40\%$ (\S\ref{sec:experiment}).

\end{itemize}

\section{Related Work}
\label{sec:related-work}

Integrating agentic reasoning into scientific orchestration has recently attracted growing attention. Most workflow-oriented efforts take a conservative approach: agents perform narrowly scoped functions while established WMS components retain authority over validation, scheduling, and execution. Pegasus adds specification-driven workflow generation, validation, debugging, submission, and monitoring~\cite{thareja2026specification}; HyperFlow limits LLM reasoning to translating scientific questions into structured intents~\cite{balis2026researchquestion}; and Parsl has been integrated with LangChain/LangGraph tool calling to enable LLM agents to execute scientific tasks on HPC resources~\cite{ma2025connecting}.
% In each case, conventional planners and execution engines remain responsible for realizing the workflow. Similarly, generative AI has been used to transform terminal sessions, notebooks, and natural-language requests into validated workflows
Related efforts use generative AI to transform terminal sessions, notebooks, or natural-language requests into executable and validated workflows~\cite{masera2025snakemaker,alam2025promptpipeline,clarke2025playbook}.
These systems demonstrate that LLMs can assist important parts of the workflow lifecycle while relying on mature WMS mechanisms to preserve reliable execution. This approach limits the consequences of incorrect agent decisions, but also confines agentic reasoning to particular stages rather than allowing it to participate continuously in orchestration.

A complementary body of work focuses on strengthening the reasoning and coordination capabilities needed for agents for broader responsibilities. Foundational prompting and reasoning techniques such as chain-of-thought prompting~\cite{wei2022cot}, ReAct~\cite{yao2023react}, Tree-of-Thoughts~\cite{yao2023tree}, and Reflexion~\cite{shinn2023reflexion} established that LLMs can interleave deliberation, tool invocation, and self-critique to produce multi-step plans. Multi-agent frameworks such as 
% MetaGPT~\cite{hong2024metagpt} and 
AutoGen~\cite{wu2024autogen} extend these capabilities to collections of role-specialized agents that coordinate through structured interaction, while AgentGen~\cite{hu2025agentgen} explores how planning capabilities can be developed and evaluated systematically at scale.
These techniques provide important building blocks for agentic orchestration, but their evaluation largely targets general-purpose reasoning, coding, or interactive environments. Scientific workflows impose additional requirements, including explicit task and data dependencies, heterogeneous computing resources, long-running execution, failures, provenance, and interaction with established workflow runtimes. Applying these reasoning capabilities to scientific orchestration therefore still requires mechanisms that connect agent reasoning to workflow state, decisions, and execution.

Other systems address this integration more directly by giving intelligent components greater authority during workflow execution. SWARM distributes functions such as job selection, scheduling, data management, and failure recovery among cooperating intelligent agents~\cite{balaprakash2025swarm}. RHAPSODY allows simulation runtimes, inference services, and agent-driven control to coexist through shared abstractions for tasks, services, resources, and execution policies~\cite{alsaadi2025rhapsody}. Academy provides abstractions for deploying and coordinating stateful agents across HPC systems, scientific instruments, and data repositories~\cite{pauloski2025academy}. FireWorks-based autonomous campaigns use machine-learning models to analyze intermediate results and prioritize subsequent calculations~\cite{takahashi2023autonomous}. These systems expose more execution state and authority to intelligent components, enabling greater automation and runtime adaptation. Yet due to the potential impact of incorrect or unexpected decisions, their agentic capabilities remain limited to particular layers or decision classes rather than controlling the complete workflow lifecycle.

Taken together, prior work demonstrates several viable ways to incorporate AI into scientific workflows, ranging from constrained assistance to direct runtime control. What remains missing is a common architectural basis for comparing and composing these choices.
Existing AI-workflow integrations remain fragmented and system-specific~\cite{shin2025revolution}, while different forms of coupling among AI models, simulations, and workflow control impose distinct middleware, execution, and performance requirements~\cite{brewer2024ai}.
Consequently, each WMS must independently determine where agents should participate, what state they should observe, which decisions they should control, how those decisions should be realized, and how their actions should be constrained and validated. Repeating these design decisions for every WMS requires substantial architectural modification and makes systematic evaluation across models, agent configurations, workflows, and infrastructure conditions difficult.

\newcolumntype{Y}{>{\raggedright\arraybackslash}X}

\begin{table*}[t]
\centering
\caption{Survey of representative scientific workflow frameworks based on how they manage scientific computations through framework-native logical abstractions, decisions over those abstractions, runtime realizations, and exposed attributes.}
\label{tab:framework-abstractions}
\footnotesize
\setlength{\tabcolsep}{0pt}
\renewcommand{\arraystretch}{1.05}

\begin{tabularx}{\textwidth}{
    @{}
    >{\raggedright\arraybackslash}p{0.09\textwidth}
    @{\hspace{6pt}}
    >{\raggedright\arraybackslash}p{0.2\textwidth}
    @{\hspace{6pt}}
    >{\raggedright\arraybackslash}p{0.2\textwidth}
    @{\hspace{6pt}}
    >{\raggedright\arraybackslash}p{0.23\textwidth}
    @{\hspace{6pt}}
    >{\raggedright\arraybackslash}p{0.26\textwidth}
    @{}
}
\toprule
\textbf{Framework} & \textbf{Logical abstractions} & \textbf{Logical decisions} & \textbf{Runtime realizations} & \textbf{Attributes} \\
\midrule

Parsl~\cite{babuji2019parsl}
&
App, task, AppFuture/DataFuture
&
Instantiate \textbf{Apps}.\newline
Resolve/release \textbf{tasks}.
&
Dispatch \textbf{tasks}.\newline
Resolve \textbf{AppFuture/DataFuture}.
&
\textbf{Task}: pending, running, failed.\newline
\textbf{AppFuture/DataFuture}: unresolved, resolved, failed.
\\
\midrule

Pegasus~\cite{deelman2015pegasus}
&
Abstract Workflow, Job, Transformation, File
&
Plan \textbf{Abstract Workflow}.\newline
Order \textbf{Jobs}.\newline
Bind \textbf{Transformations}.
&
Map/submit \textbf{Jobs}.\newline
Stage/register \textbf{Files}.
&
\textbf{Job}: planned, running, failed.\newline
\textbf{File}: unstaged, staged, registered.
\\
\midrule

Cromwell~\cite{cromwell2026}
&
WOM graph, WOM node, Value Store
&
Expand \textbf{WOM graph}.\newline
Resolve/release \textbf{WOM nodes}.
&
Execute \textbf{WOM nodes}.\newline
Populate \textbf{Value Store}.
&
\textbf{WOM node}: not started, running, done.\newline
\textbf{Value Store}: unavailable, available.
\\
\midrule

Toil~\cite{vivian2017toil}
&
Job graph, Job, Promise
&
Extend \textbf{Job graph}.\newline
Resolve/order \textbf{Jobs}.
&
Submit/execute \textbf{Jobs}.\newline
Resolve \textbf{Promises}.
&
\textbf{Job}: waiting, running, completed.\newline
\textbf{Promise}: unresolved, resolved.
\\
\midrule

CWL~\cite{crusoe2022cwl}
&
Workflow, WorkflowStep, Process
&
Validate \textbf{Workflow}.\newline
Resolve \textbf{WorkflowSteps}.\newline
Bind \textbf{Process} inputs.
&
Instantiate \textbf{Processes}.\newline
Execute \textbf{Processes} through runners.
&
\textbf{WorkflowStep}: inputs, outputs, scatter.\newline
\textbf{Process}: inputs, outputs, requirements.
\\
\midrule

WDL~\cite{openwdl2026}
&
workflow, task, call
&
Expand \textbf{workflow}.\newline
Resolve \textbf{calls}.\newline
Bind \textbf{call} inputs.
&
Construct \textbf{task} commands.\newline
Execute \textbf{calls}.
&
\textbf{Task}: inputs, outputs, runtime requirements.\newline
\textbf{Call}: inputs, alias, dependencies.
\\
\midrule

Snakemake~\cite{koster2012snakemake}
&
Snakefile, rule, wildcard
&
Parse \textbf{Snakefile}.\newline
Resolve \textbf{rules}.\newline
Bind \textbf{wildcards}.
&
Execute \textbf{rules}.\newline
Materialize \textbf{rule} outputs.
&
\textbf{Rule}: input, output, resources.\newline
\textbf{Wildcard}: value, constraints.
\\
\midrule

Makeflow~\cite{albrecht2012makeflow}
&
DAG file, rule, source/target file
&
Parse \textbf{DAG file}.\newline
Resolve/release \textbf{rules}.
&
Execute \textbf{rules}.\newline
Transfer \textbf{source/target files}.
&
\textbf{Rule}: sources, targets, command.\newline
\textbf{File}: missing, available, outdated.
\\
\midrule

Nextflow~\cite{ditommaso2017nextflow}
&
process, channel, dataflow value
&
Trigger \textbf{processes}.\newline
Bind \textbf{channels/dataflow values}.
&
Execute \textbf{processes}.\newline
Stage/cache process data.
&
\textbf{Process}: inputs, outputs, directives.\newline
\textbf{Channel/dataflow value}: waiting, emitted, closed.
\\
\midrule

FireWorks~\cite{jain2015fireworks}
&
Workflow, Firework, Firetask
&
Update \textbf{Workflow}.\newline
Prioritize \textbf{Fireworks}.
&
Reserve \textbf{Fireworks}.\newline
Execute \textbf{Firetasks}.
&
\textbf{Firework}: ready, running, completed.\newline
\textbf{Workflow}: ready, running, completed. 
\\

\bottomrule
\end{tabularx}
\end{table*}

\section{Avatar: Reference Architecture}
\label{sec:reference-arch}

\subsection{Deriving the Reference Architecture}
\label{sec:reference-arch-formulation}

The fragmented and system-specific integrations identified above motivate a common architectural basis for introducing agentic reasoning into scientific workflows. Rather than deciding independently for each WMS where an agent should be placed and what it should control, we derive a \textit{reference architecture} from responsibilities that recur across existing WMSs.
\autoref{tab:framework-abstractions} summarizes our survey of representative WMSs. Despite substantial differences in architecture, programming model, execution environment, and implementation, these systems exhibit a common \textit{abstraction-oriented design pattern}.
They represent scientific computations through \textit{logical abstractions}, such as workflows, tasks, data objects, and dependencies; make \textit{logical decisions} over these abstractions to determine \emph{what} should happen; and realize those decisions through \textit{runtime} mechanisms that determine \textit{how} the computation is executed. This separation between logical workflow management and runtime realization provides the first foundation for Avatar.

The survey further shows that workflow abstractions are accompanied by \textit{attributes} describing either their intended configuration or their observed runtime state. 
Examples include task readiness, priority, resource requirements, execution status, data availability, retry count, and failure state. 
These attributes serve as the interface between workflow specification and execution by exposing decision points throughout the workflow lifecycle. For instance, newly available inputs may enable downstream tasks, execution failures may trigger retries, and resource unavailability may necessitate replanning. Consequently, WMSs must not only represent the intended computation, but also observe how that computation evolves and preserve enough execution history to support subsequent decisions.

\begin{figure}[t]
    \centering
    \includegraphics[width=0.98\linewidth,trim={70 160 90 180}, clip]{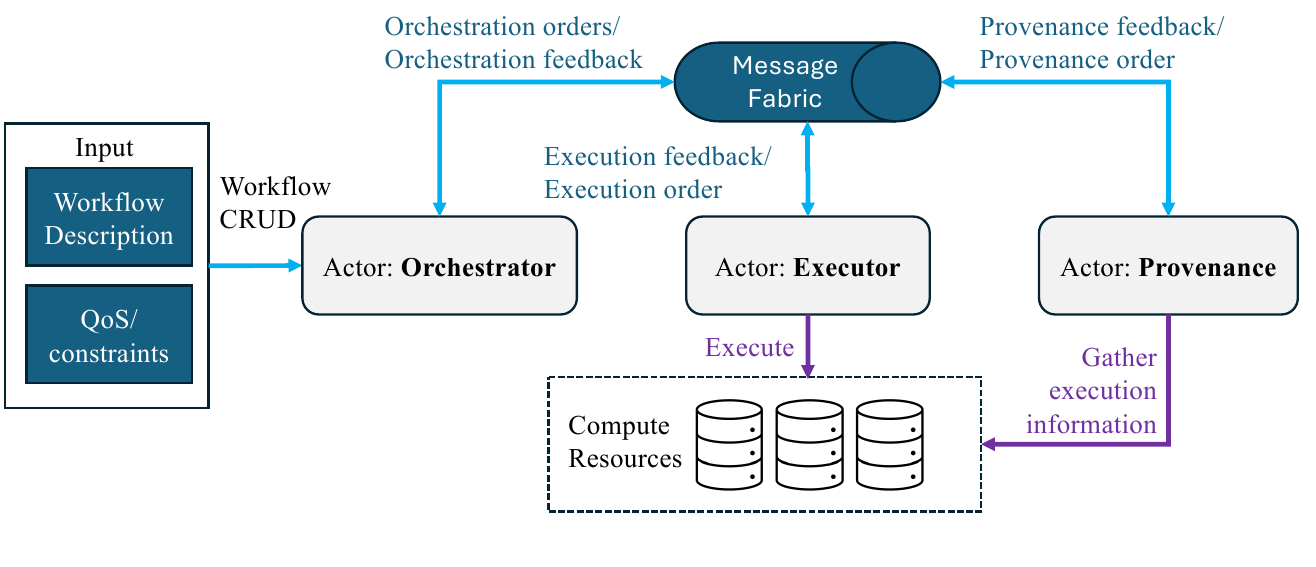}
    \caption{Avatar architecture}
    \vspace{-0.25in}
    \label{fig:ref-arch}
\end{figure}

Together, these recurring patterns yield three fundamental responsibilities for scientific workflow orchestration. First, a system must maintain the logical representation of a computation and decide how that representation should evolve. Second, it must realize those decisions through concrete execution mechanisms. Third, it must observe and preserve the evidence required to determine when decisions should be made, validated, or reconsidered. Avatar maps these responsibilities to three corresponding components, shown in \autoref{fig:ref-arch}. The \textit{orchestrator} manages the logical abstractions and their evolution. It interprets workflow descriptions, user intent, QoS requirements, execution constraints, and makes logical decisions concerning workflow construction, task ordering, priorities, adaptation, and recovery. The \textit{executor} realizes these decisions on concrete infrastructure. It selects and controls execution mechanisms, binds tasks to resources, launches tasks, stages data, manages runtime execution, and reports outcomes. The \textit{provenance} component maintains the evidence needed to close the decision loop. It determines which attributes and events should be collected, records logs, performance measurements, task outcomes, failures, retries, and workflow history, and emits triggers when observed outputs require a new decision.

These components describe normalized responsibilities rather than modules that existing WMSs necessarily expose under the same names. Pegasus, for example, separates an abstract workflow from the executable workflow generated for a target environment whereas Parsl separates dynamic dependency management from execution through configurable executors. Rule- and dataflow-oriented systems such as Snakemake, Makeflow, and Nextflow use different abstractions, yet follow the same progression from logical dependency reasoning to runtime realization
\cite{koster2012snakemake,albrecht2012makeflow,ditommaso2017nextflow}. Systems such as Cromwell, Toil, and FireWorks further demonstrate the importance of preserving execution state and history for monitoring, retry, recovery, and dynamic workflow modification
\cite{cromwell2026,vivian2017toil,jain2015fireworks}.

This decomposition directly addresses the architectural challenge identified earlier by making orchestration responsibilities explicit and allowing the decision mechanism within each responsibility to vary independently. Rather than embedding agents at system-specific locations, Avatar lets the orchestrator, executor, and provenance components use conventional algorithms, LLM-backed agents, or hybrid policies while preserving common interfaces and validation boundaries. This makes Avatar \textit{representative} of recurring WMS responsibilities yet \textit{general} enough to support different agentic configurations, providing a common basis for locating, composing, and systematically evaluating where agentic reasoning is introduced and what trade-offs it creates.

\subsection{Actor-based Realization}
\label{sec:reference-arch-actor}

\begin{figure}
    \centering
    \includegraphics[width=0.95\linewidth, trim={140 200 110 200}, clip]{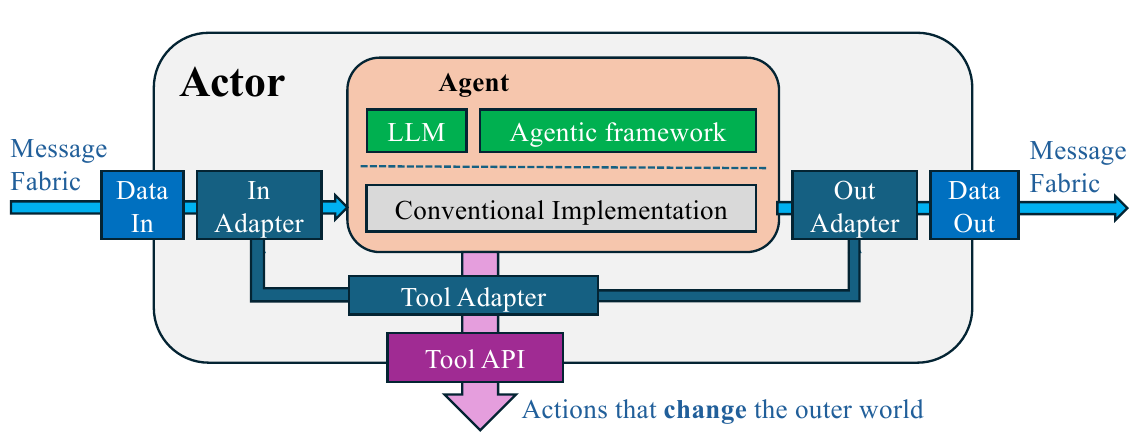}
    \caption{Actors integrated into Avatar}
    \vspace{-0.25in}
    \label{fig:ref-arch-actor}
\end{figure}

We realized the core components (i.e., the orchestrator, executor, and provenance) as \textit{actors} with a standardized interface for communication and action (\autoref{fig:ref-arch-actor}). An actor's internal decision maker, referred to as its \textit{agent}, may be implemented using conventional algorithms, LLM-backed reasoning, or a hybrid of the two. This abstraction enables existing planners, executors, and provenance managers to be wrapped and systematically compared with agentic implementations without changing the surrounding architecture. Actors interact through a shared \textit{message fabric}, following the message-oriented communication model common in distributed systems. The message fabric decouples components, makes interactions explicit, and provides a controlled interface for information exchange, allowing message schemas, routing policies, and access controls to be varied without modifying component internals or the overall architecture.

Each actor interacts with the rest of the system through three controlled interfaces: data input, data output, and tool invocation. The data input and output interfaces connect the actor to the message fabric. The tool interface allows the actor to affect the external environment by invoking predefined tools, such as workflow submission APIs, resource managers, monitoring services, provenance stores, or MCP servers. Before messages reach the agent, they pass through \textit{adapters} that translate external data into the representation expected by the actor. Agent outputs also pass through adapters before they are exposed to the rest of the system.
Adapters provide both portability and safety. They allow existing workflow components to be wrapped as actors, or replaced with LLM-backed agents, without changing the rest of the architecture. They also serve as policy-enforcement boundaries: instead of giving agents unrestricted access to the runtime, adapters control what information reaches the agent and what actions can leave it. For example, adapters can filter runtime state, normalize provenance records, validate action formats, reject unsafe tool calls, or require additional checks before executing an agent-generated decision.

This design makes the study of agentic scientific workflow orchestration explicit and systematic. By varying actor implementations, exposed state, available tools, and message-fabric policies, we can compare fully conventional baselines, partially agentic systems, and more autonomous orchestration loops before committing to a system-specific integration. 
By making observation, reasoning, action, and validation boundaries explicit, Avatar provides both a practical blueprint for agentic workflow orchestration and an experimental framework for understanding its benefits and limitations. 

%========================================================
\section{Implementation}\label{sec:implementation}

Avatar is implemented in Python~3.12 on \emph{Academy}~0.5.0, a middleware for stateful agents on federated research infrastructure~\cite{pauloski2025academy}. Academy directly supplies the primitives that Avatar's actor model requires. An actor is written as an Academy \texttt{Agent} whose \texttt{@action}-decorated methods expose the operations that peers may invoke remotely, and whose \texttt{@loop}-decorated methods run its autonomous control loop. Actors reference one another through \texttt{Handles} and exchange messages asynchronously through per-agent mailboxes routed by an \texttt{Exchange}, which serves as our message fabric. 
In the current version, Avatar uses \texttt{LocalExchangeFactory} (an in-process mailbox exchange) backed by a \texttt{ThreadPoolExecutor}. Task execution uses Parsl~2026.3.9 and native CCTools~TaskVine; the Colmena surrogate is PyTorch (CUDA~12.2) over RDKit fingerprints; agent reasoning calls use OpenAI client~2.28.0. 

\textbf{Actors}. The \emph{provenance} agent (\texttt{ProvenanceActor}) ingests the executor's \texttt{TaskEvent} stream, keeps per-task attempt and lineage state, and on each failure emits a \texttt{Trigger} carrying a diagnosis produced by its policy. The \emph{executor} (\texttt{ExecutorActor}) owns a bounded worker pool and dispatches each task through one overridable interface. \texttt{\_execute} is a local \textit{job.py} subprocess for TaskVine and \texttt{python\_app} for Parsl, reporting outcomes and events back to provenance and realizing \texttt{scale-out}/\texttt{scale-in} on the resource pool. The \emph{orchestrator} (\texttt{OrchestratorActor}) runs the control loop: it releases dependency-ready tasks, drains completions and failure triggers, maps each diagnosis to a recovery \texttt{Decision}, and enacts it once validated. 

\textbf{Policies, adapters, and modes.} 
Each agent's decision-maker is a pluggable \emph{policy} which is a deterministic rule set derived from the native backend or an LLM-backed reasoner. 
Every proposed action passes an \emph{adapter} that validates it against one fixed catalog: \textit{retry}, \textit{migrate}, \textit{blacklist-worker}, \textit{replicate-task}, \textit{scale-out}/\textit{in}, \textit{batch}, \textit{throttle}, and the steering actions \textit{propose-batch}, \textit{retrain-surrogate}, \textit{infer-pool}, and \textit{stop-campaign}, so agentic and rule modes share one action set and invalid actions are counted. This yields three variants of implementation: \textbf{M0} uses rule policies throughout; \textbf{M1} adds an LLM diagnosis in provenance while rules still act; \textbf{M2} lets the LLM both diagnose and decide in the orchestrator, with the executor still actuating by rule.

An LLM-backed \emph{executor}, i.e., agentic \emph{placement} across heterogeneous resources, is a natural third locus for reasoning, but our single-node, homogeneous testbeds in this paper present no real placement choice, so we leave it to the multi-node setting for future.

\section{Evaluation}\label{sec:experiment}

\subsection{Methodology}

\medskip
\noindent\textbf{Objectives.} We assess whether Avatar provides an appropriate reference architecture for studying agentic integration into scientific workflow systems along three dimensions. (i) \textit{Representativeness}: whether Avatar can reproduce the control behavior and outcomes of distinct conventional WMSs while preserving the same core actor structure. (ii) \textit{Generalizability}: whether Avatar can support workflows that differ substantially in structure, execution pattern, and scientific domain without redesigning its core components. (iii) \textit{Applicability}: whether Avatar can serve as an experimental instrument for identifying where, when, and how agentic reasoning is beneficial. 
\medskip
\noindent\textbf{Metrics.} We align the evaluation criteria with these objectives. \textit{Representativeness} is assessed through workflow correctness and behavioral agreement with native WMS implementations, including equivalent completed work and control outcomes. \textit{Generalizability} is demonstrated by executing all diverse workloads with the same Avatar actors and interfaces. For \textit{applicability}, we measure workload-specific benefits and costs: retries and wasted worker time for resilience, backlog and resource utilization for elastic scaling, and scientific outcome, molecule evaluations, GPU usage for molecular design.

\medskip
\noindent\textbf{Workloads.}
We evaluate Avatar on three workloads chosen to span distinct workflow organizations, orchestration structures, and objectives.
\begin{itemize}[topsep=0pt,itemsep=0pt,leftmargin=*]
    \item \textit{Resilience (E1):} A fault-prone fan-out/fan-in DAG representative of traditional scientific workflows: one input task feeds $N$ parallel map tasks followed by a reduce task. Each map task performs $300$ iterations of a normalized $160\times160$ double-precision matrix multiplication ($a \leftarrow (a\cdot a)/||a||$), but its execution is intentionally unreliable. After the computation, with probability $p$, a task either fails or becomes a straggler by sleeping for an additional $0.15$s. Failures are either \textit{transient}, disappearing after a retry, or \textit{permanent}, recurring across all retries, randomly determined by an experiment parameter $x$ which indicates the statistical fraction of permanent failures. By default, we set $p=0.35$ and $x=1/3$. On failure, the task exits with a non-zero status and exposes only the observed error symptom to the recovery policy; each attempt is also recorded in a shared log with its true class for offline evaluation. Avatar executes the tasks as local \texttt{asyncio} subprocesses driven by its three Academy actors through a \textit{LocalExchangeFactory}, under the same retry ceiling as the native baseline. Its \textit{Provenance} actor diagnoses failures from their retry history.
    
    \item \textit{Scaling (E2):}
    A dynamic analyze-and-steer streaming workflow in which $N$ inputs arrive over time, each triggering a three-stage sub-workflow: \textit{preprocess} $\rightarrow$ \textit{analyze} ($\times K$) $\rightarrow$ \textit{aggregate}. We set $K=4$, so each input creates four parallel analysis tasks. Each task performs a dense matrix multiplication similar to E1 and starts only after all tasks in the preceding stage have completed. Inputs follow a non-stationary schedule with four phases: (i) low ($\approx2f$/s for 4s), (ii) burst ($\approx15f$/s for 5s), (iii) high ($\approx8f$/s for 5s), and (iv) recover ($\approx2f$/s for 4s) where $f \approx N/131$ is the rate control parameter at a given $N$, selected to make the input arrival duration finishes in $\approx18$s.

    \item \textit{Active Learning (E3):} An unmodified Colmena campaign that maximizes the HOMO--LUMO \texttt{gap} over a pool of QM9 molecules. The candidate pool contains $n_p=5000$ QM9 molecules together with an initial labeled seed set of $n_s=100$. In each round, the campaign retrains an MLP surrogate over Morgan fingerprints on the GPU using all molecules scored so far, runs inference over the candidate pool, greedily selects the top-$B$ candidates ($B=32$) by predicted gap, and evaluates the selected batch. The evaluation uses a deterministic QM9-\texttt{gap} lookup as a stand-in for an expensive property calculation. Avatar drives the campaign through its three Academy actors, with every steering proposal (i.e., \textit{propose-batch}, \textit{retrain-surrogate}, \textit{infer-pool}, and \textit{stop-campaign}) expressed as a catalog action and validated by an adapter before execution.
\end{itemize}

\medskip
\noindent\textbf{Baselines.} In addition to Avatar, each workload is paired with the conventional mechanism used to manage the corresponding execution in practice.

\begin{itemize}[topsep=0pt,itemsep=0pt,leftmargin=*]
\item \textit{TaskVine.} For E1, TaskVine drives the resilience workload on the compute node using its co-located \texttt{vine\_worker} and blind fixed retries with an upper bound $R$.

\item \textit{Parsl.} For E2, each scaling-workload task is implemented as a Parsl \texttt{python\_app} and executed using fixed Parsl pools of different sizes: \texttt{w4} (small), \texttt{w12} (medium), and \texttt{w32} (large). Each pool provides concurrency up to its worker count, and backlog is sampled every $0.2$s.

\item \textit{Colmena~\cite{ward2021colmena}.} For E3, the native controller of the E3 Active-learning workload that use a \textit{BaseThinker} over a \textit{ParslTaskServer} with three registered task methods (\textit{evaluate\_molecule}/\textit{train\_model}/\textit{infer\_model}) on a GPU Parsl executor, running a fixed round budget of $12$.
\end{itemize}

\medskip
\noindent\textbf{System Configurations.}
All experiments run on the University of Chicago RCC \textit{midway3} cluster. For E1, Avatar is submitted as an \texttt{sbatch} job running in \textit{exclusive} mode on a single \texttt{caslake} compute node with an Intel(R) Xeon(R) Gold 6248R CPU at 3.00\,GHz, $48$ CPU cores, and $192$\,GB of RAM. The native TaskVine Manager, its co-located \texttt{vine\_worker}, and the \textit{job.py} tasks execute on the same node, while Avatar's three Academy actors run in-process alongside them.
For E2, Avatar launches its agents on the login node (\texttt{midway3-login3}), while Parsl's \textit{SlurmProvider(exclusive=True)} provisions a dedicated \textit{caslake} compute node with the same hardware specification as E1 for the numerical workers. Exclusive allocation ensures that Parsl workers receive dedicated CPU cores and avoids interference from co-located workloads.
For E3, the Colmena steering workload is submitted as an \texttt{sbatch} job to the \texttt{gpu} partition with $1$ GPU and \texttt{--cpus-per-task=16}, with the Academy actors running alongside the Parsl workers. The GPU is an NVIDIA Quadro RTX~6000 running CUDA~12.2. Across all experiments, LLM API calls from Avatar actors use the ALCF Sophia inference endpoint (\texttt{inference-api.alcf.anl.gov}) and the \texttt{open-ai/gpt-oss-20b} model hosted at Argonne National Laboratory.

\begin{figure}[t]
    \centering
    \subfloat[Reproducibility on E1 (solid = productive, hatched = wasted; whiskers = min--max over 3 runs).]{\includegraphics[width=0.46\linewidth]{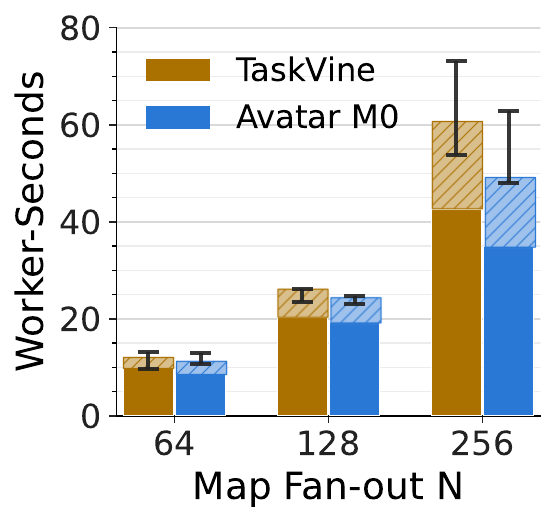} \label{fig:e1-representative}}
    ~~
    \subfloat[Reproducibility on E2]{\includegraphics[width=0.46\linewidth]{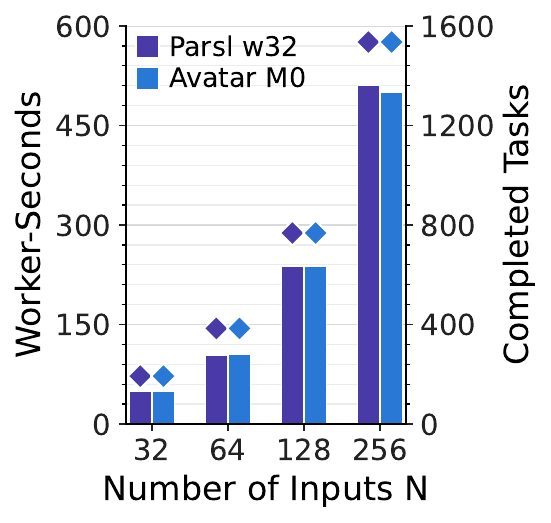}\label{fig:e2-representative}} 
    \caption{Avatar is representative and generalizable across WMSs}
    \vspace{-0.1in}
    \label{fig:representative}
\end{figure}

\subsection{Experimental Results} 

\medskip
\noindent\textbf{Representativeness.}
To evaluate Avatar's \textit{representativeness}, we use its rule-based configuration (M0), which introduces no agentic decision making. If Avatar captures the essential orchestration behavior of existing workflow systems, M0 should reproduce their characteristic behavior under equivalent workloads and policies.

We compare Avatar M0 with native TaskVine on E1 (Resilience) under identical scale ($N$) and fault plans using the same failure-probability seeds and retry ceiling $R{=}2$. Avatar M0 reproduces TaskVine's recovery behavior \emph{exactly}: retry, completion, and permanent-failure counts match for every $(N,\textit{seed})$ pair (e.g., both produce $88$ retries, $231$ completions, and $26$ permanent failures at $N{=}256$). Fig.~\ref{fig:e1-representative} further shows that both systems exhibit the same scaling trend in used and wasted worker-seconds, with differences remaining within the variation across seeds. Avatar M0 consistently consumes slightly fewer worker-seconds because its simplified executor avoids part of the dispatch overhead incurred by TaskVine's full \texttt{vine\_worker} implementation. This difference becomes more visible as fan-out increases, reaching approximately $19\%$ at $N{=}256$. Importantly, this implementation-level difference does not change the workflow's recovery decisions or outcomes. Avatar therefore preserves the behavior relevant to the experiment.

\vspace{2mm}
\noindent
\colorbox{blue!10}{
\parbox{0.96\linewidth}{
\underline{\textbf{Takeaway (Representativeness)}:} \emph{Avatar reproduces native workflow behavior across distinct systems through a representative, lightweight architecture.
}}}

\medskip
\noindent\textbf{Generalizability.}
We continue with M0 to evaluate Avatar's \textit{generalizability}. If Avatar is generalizable, the same core architecture should support qualitatively different workflow structures and execution models without redesigning its actors or message fabric.

\begin{figure}[t]
    \centering
    \includegraphics[width=0.45\linewidth]{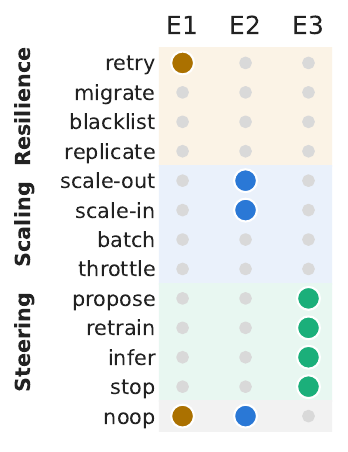}
    \caption{
    One shared action catalog spans three workloads on one unchanged core. Filled = invoked; faint = in catalog but not triggered.}
    \vspace{-0.1in}
    \label{fig:generative}
\end{figure}

Fig.~\ref{fig:e2-representative} provides the first evidence by complementing the E1 result with a structurally different workload. Across increasing workload intensities $N$, Avatar M0 and native Parsl complete the same number of tasks, while their consumed worker-seconds differ by less than $2\%$. Together with E1, this shows that the same Avatar implementation remains generalizable in two contrasting settings: E1 is a batch-oriented, fault-prone DAG with relatively stable computational demand, whereas E2 is a streaming workflow with bursty arrivals and dynamically changing resource demand. Their native implementations also rely on different workflow systems, TaskVine and Parsl, which expose different integration capabilities with Avatar's core components (Table~\ref{tab:framework-abstractions}). Avatar's ability to reproduce both therefore demonstrates that its architecture is not tied to the execution model or integration characteristics of a particular WMS.

Fig.~\ref{fig:generative} provides stronger implementation-level evidence. Across E1--E3, Avatar uses a single shared action catalog, organized into resilience, scaling, and scientific-steering capabilities. The $\approx1200$ lines implementing the core orchestrator, executor, and provenance actors, together with their policies, adapters, action catalog, and message types, remain byte-identical across all three workloads. Each workload connects to this core only through a one-method \texttt{\_execute} interface: a \textit{job.py} subprocess for E1 and a Parsl \texttt{python\_app} for E2 and E3, together with its workload generator. E3 additionally introduces only a small orchestrator subclass for extracting application-specific results. 
Despite sharing the same core, the workloads exercise largely disjoint parts of the action catalog: E1 uses resilience actions, E2 uses scaling actions, and E3 uses scientific-steering actions, with \texttt{noop} as the only shared action. 
This shows that one unchanged Avatar core can drive a static fault-tolerant DAG, a dynamic elastic stream, and a GPU-based scientific steering campaign.

\vspace{2mm}
\noindent
\colorbox{blue!10}{
\parbox{0.96\linewidth}{
\underline{\textbf{Takeaway (Generalizability)}:} \emph{Avatar generalizes across diverse workflows and control objectives using one unchanged core architecture and interfaces.
}}}

\begin{figure}[t]
    \centering
    \subfloat[Varying R]{\includegraphics[width=0.35\linewidth]{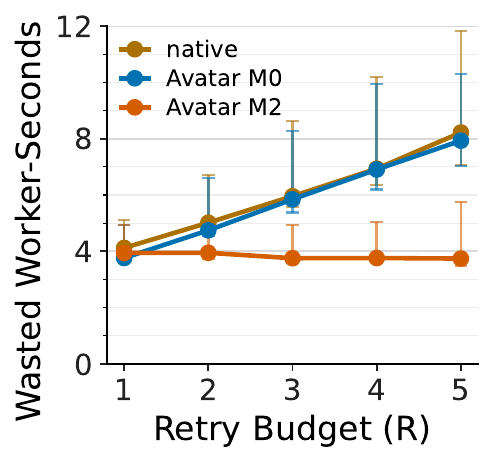}\label{fig:e1-rounds}}
    ~
    \subfloat[Varying p]{\includegraphics[width=0.32\linewidth]{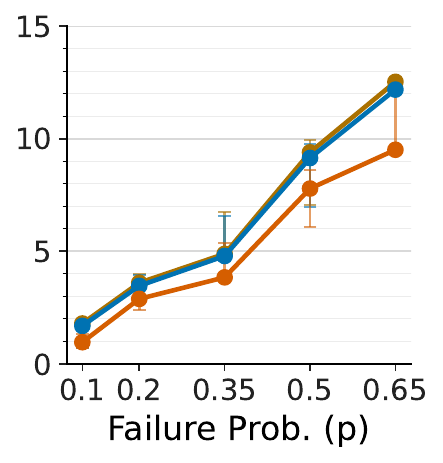}\label{fig:e1-p}}
    ~
    \subfloat[Varying x]{\includegraphics[width=0.32\linewidth]{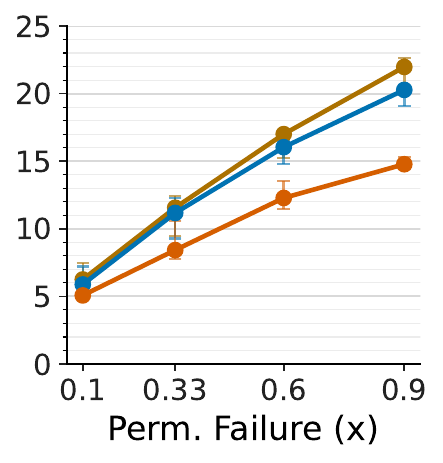}\label{fig:e1-x}}
    \caption{\textbf{E1.} Resilience on TaskVine backend ($N{=}128$).}
    \vspace{-0.25in}
    \label{fig:e1-resilience}
\end{figure}

\begin{figure}[t]
    \centering
    \subfloat[\textbf{E2.} Parsl QoS at $N{=}128$]{
    \includegraphics[width=0.43\linewidth]{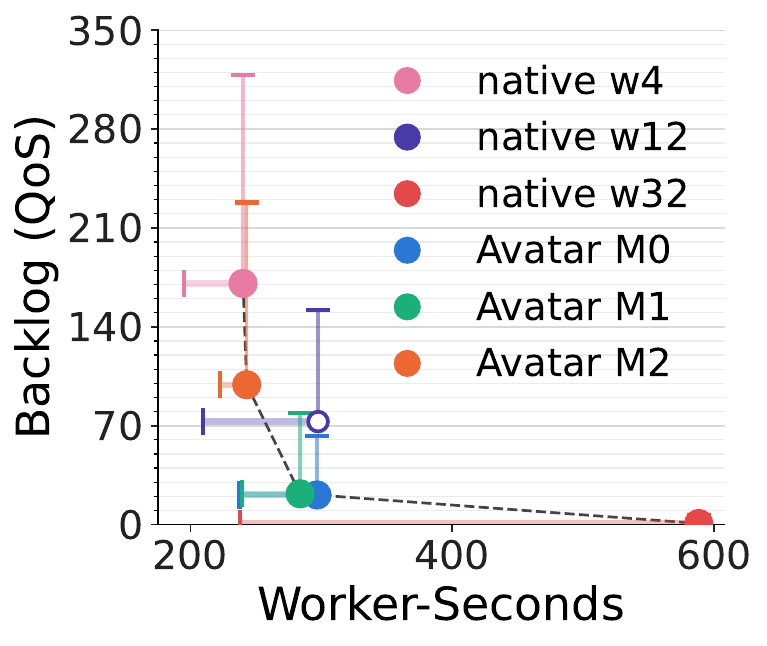}\label{fig:e2-parsl}}
    \subfloat[\textbf{E3.} Colmena molecular design]{\includegraphics[width=0.55\linewidth]{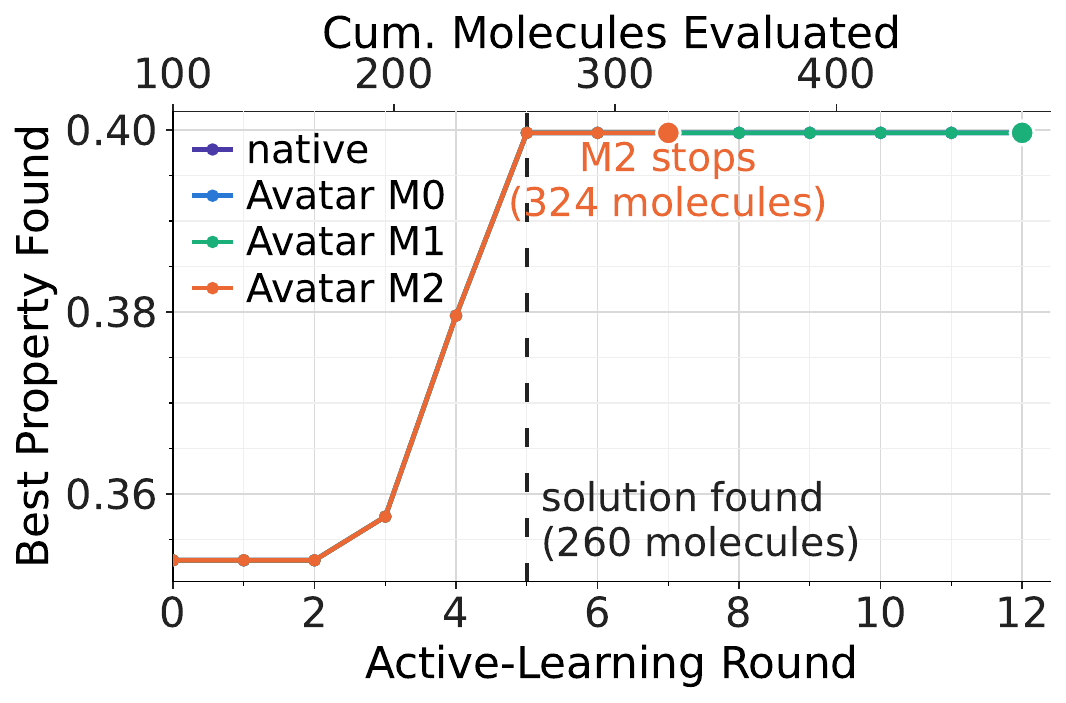}\label{fig:colmena}}
    \caption{E2 and E3 on Parsl backend.}
    \vspace{-0.25in}
    \label{fig:e2e3}
\end{figure}

\medskip
\noindent\textbf{Applicability.}
Next, we use Avatar M1 and M2 to evaluate its \textit{applicability} as an experimental instrument for studying agentic workflow control. 
A common motivation for agentic workflow management is that such dynamic reasoning can outperform conventional static policies by adapting actions to runtime conditions. 
We therefore use Avatar to test this hypothesis directly: \textit{does LLM-enabled dynamic control improve efficiency over conventional rule-based control, and under what conditions?}

On E1 (Resilience), Fig.~\ref{fig:e1-resilience} examines three factors independently at $N{=}128$: retry budget $R$ (Fig.~\ref{fig:e1-rounds}), task failure probability $p$ (Fig.~\ref{fig:e1-p}), and permanent-failure fraction $x$ (Fig.~\ref{fig:e1-x}). Native blind retry and M0 exhibit the same behavior, with wasted computation increasing as each factor grows. M2 instead uses the observed failure history to identify permanently failing tasks and gives up after one retry. Consequently, its wasted computation is effectively \emph{$R$-invariant}, while its advantage over blind retry increases as failures become more frequent ($p$) or more likely to be permanent ($x$). The experiment therefore reveals a clear regime in which agentic reasoning is valuable: diagnosis-driven recovery provides the greatest benefit when failures are both frequent and unrecoverable.

E2 exposes the opposite regime. Fig.~\ref{fig:e2-parsl} shows the cost--QoS trade-off for Parsl scaling at $N{=}128$. Among the fixed pools, \texttt{w4} minimizes resource allocation but accumulates substantial backlog (mean ${\approx}171$), whereas \texttt{w32} maintains near-zero backlog by provisioning approximately $2.5\times$ the compute it actually uses. Avatar's rule-based autoscaler (M0/M1) approaches the desirable low-backlog, low-waste region with a mean backlog of approximately $21$. In contrast, M2 reaches a mean backlog of approximately $99$ because the LLM cannot react as quickly as the scaling loop requires. Thus, dynamic agentic reasoning is not universally advantageous: for high-frequency, latency-sensitive control, the reasoning latency itself becomes a control cost, and a lightweight rule can be more effective.

Finally, E3 demonstrates where dynamic agentic control can provide substantial benefit in a real scientific steering application. Fig.~\ref{fig:colmena} relates the Colmena learning curve to GPU cost. The native baseline and M0/M1 execute the full $12$-round campaign, whereas M2 reasons over the observed learning progress, detects convergence, and terminates at round $7$. It reaches the \emph{same} best HOMO--LUMO gap ($0.3997$) while reducing GPU-busy time by approximately $40\%$. Unlike E2's fast scaling loop, convergence detection is an infrequent, context-dependent decision for which reasoning latency is small relative to the computation it can avoid.

Taken together, these experiments refine the original hypothesis: agentic reasoning is most beneficial for relatively infrequent decisions that require interpreting execution context and can avoid substantial downstream work, while conventional rules remain preferable for high-frequency control. 
More importantly, Avatar makes this distinction experimentally observable by allowing the decision policy to change while holding the surrounding architecture fixed.

\vspace{2mm}
\noindent
\colorbox{blue!10}{
\parbox{0.96\linewidth}{
\underline{\textbf{Takeaway (Applicability)}:} \emph{Avatar demonstrates practical applicability by enabling controlled, comparable evaluation of agentic policies across diverse workflow scenarios.
}}}

\section{Conclusion and Future Work}\label{sec:conclusion}
In this paper, we propose Avatar, which recasts a scientific WMS using three cooperating actors: orchestrator, executor, and provenance, with pluggable decision policies. 
This allows the same implementation to support both deterministic rules and LLM-based reasoning. 
Our experiments show that Avatar is \emph{representative}, reproducing native TaskVine and Parsl execution in rule mode; \emph{generalizable}, supporting static fault-tolerant DAGs, dynamic streaming workflows, and GPU-based active learning; and \emph{applicable}, with LLM policies providing benefits for decisions that require inference.

We plan to extend Avatar to multi-node and heterogeneous environments spanning CPUs, GPUs, and multiple sites. This will introduce resource-selection decisions for the executor, such as selecting a resource for a task or migrating a failed task to another node. This setting will allow us to extend LLM-based decision-making to the executor and evaluate the safety of the resulting action space. We also plan to improve failure handling by identifying error types and applying targeted recovery actions rather than simply retrying failed tasks.

\balance
\bibliographystyle{IEEEtran}
\bibliography{references}

\end{document}